\pdfoutput=1
\documentclass[manuscript,nonacm]{acmart}
\usepackage{todonotes}

\usepackage{multirow}
\usepackage{tabls}
\usepackage{wrapfig}
\usepackage{subfigure}
\usepackage{listings}
\usepackage{amsmath,amsthm}
\usepackage{etoolbox}
\usepackage{array}
\usepackage{tikz}
\usetikzlibrary{shapes,arrows,arrows.meta, positioning,shapes.misc,decorations.markings}
\usepackage{adjustbox}
\usepackage{hyperref}
\usepackage{booktabs}
\usepackage{lipsum,multicol}
\usepackage[font=small,skip=4pt]{caption}
\usepackage{enumitem}
\usepackage{caption}

\setcopyright{acmcopyright}
\copyrightyear{2023}
\acmYear{2023}
\acmDOI{XXXXXXX.XXXXXXX}

\author{Robert J. Joyce}
\affiliation{%
   \institution{Booz Allen Hamilton}
   \institution{University of Maryland, Baltimore County}
   \country{USA}
}
\email{joyce_robert2@bah.com}

\author{Tirth Patel}
\affiliation{%
   \institution{University of Maryland, Baltimore County}
   \country{USA}
}

\email{tpatel9@umbc.edu}

\author{Charles Nicholas}
\affiliation{%
   \institution{University of Maryland, Baltimore County}
   \country{USA}
}
\email{nicholas@umbc.edu}

\author{Edward Raff}
\affiliation{%
   \institution{Booz Allen Hamilton}
   \institution{University of Maryland, Baltimore County}
   \country{USA}
}
\email{raff_edward@bah.com}

\begin{document}

\title{AVScan2Vec: Feature Learning on Antivirus Scan Data for Production-Scale Malware Corpora}



\begin{abstract}
When investigating a malicious file, searching for related files is a common task that malware analysts must perform. Given that production malware corpora may contain over a billion files and consume petabytes of storage, many feature extraction and similarity search approaches are computationally infeasible. Our work explores the potential of antivirus (AV) scan data as a scalable source of features for malware. This is possible because AV scan reports are widely available through services such as VirusTotal and are $\approx$100$\times$ smaller than the average malware sample. The information within an AV scan report is abundant with information and can indicate a malicious file's family, behavior, target operating system, and many other characteristics. We introduce AVScan2Vec, a language model trained to comprehend the semantics of AV scan data. AVScan2Vec ingests AV scan data for a malicious file and outputs a meaningful vector representation. AVScan2Vec vectors are $\approx$3 to 85$\times$ smaller than popular alternatives in use today, enabling faster vector comparisons and lower memory usage. By incorporating Dynamic Continuous Indexing, we show that nearest-neighbor queries on AVScan2Vec vectors can scale to even the largest malware production datasets. We also demonstrate that AVScan2Vec vectors are superior to other leading malware feature vector representations across nearly all classification, clustering, and nearest-neighbor lookup algorithms that we evaluated.
\end{abstract}

\settopmatter{printacmref=false}
\maketitle
\thispagestyle{empty}

\section{Introduction}
\label{sec:introduction}

During the course of analyzing a malicious file, it is common for analysts to search a large repository of malware in an attempt to identify other related malicious files. These search queries are relied upon to track how a family of malware is evolving over time, to identify infrastructure used in a malicious campaign, and to search for prior analysis performed on similar files. Hundreds of thousands of unique, previously-unseen malicious files are observed on a daily basis \cite{virustotal}. As a result, production malware corpora may contain over a billion files, requiring petabytes of storage space and making the scope of these queries massive \cite{shadowserver}. Machine learning tasks such as classification and clustering are also increasingly relied upon to provide automation in a field where human effort is slow and costly. The efficacy of related malware queries, and of these ML tasks, are extremely dependent on the features selected to represent malware. Feature engineering of malicious files is especially challenging due to the various obfuscation and evasion techniques adversaries use to hide the true nature of their malware \cite{Raff2020a}. Prior work has popularized feature selection from a variety of sources, including raw file bytes, file format metadata, static analysis, and dynamic analysis \cite{ucci}. However, the feature extraction techniques which best counteract the adversarial nature of this problem space (i.e. disassembly and dynamic analysis) also require the most prolonged analysis and would not feasibly scale to the necessary dataset sizes. More scalable feature extraction methods are hindered by obfuscation, restricted to a single file format, and/or limited in their capacity to identify higher-level features of malware.

We recognize that antivirus (AV) scan data provides an abundance of useful meta-information about malware, yet we are not aware of any prior work that has leveraged it as a source of ML features. We present AVScan2Vec, a language model which uses self-supervised feature learning to produce embedded representations for AV scan data. Counterparts such as Word2Vec \cite{word2vec}, Doc2Vec \cite{doc2vec}, Node2Vec \cite{node2vec}, and Code2Vec \cite{code2vec} have enabled various types of data to be embedded and subsequently used for downstream ML tasks. Likewise, AVScan2Vec learns to represent an entire AV scan report as a single vector. AVScan2Vec is capable of embedding malware from the same family or with similar behavior into nearby vectors, making it conducive to nearest-neighbor lookup, clustering, and classification. Our results show that AVScan2Vec vectors outperform prior approaches on all these tasks, in some cases by an appreciable margin.

We envision AVScan2Vec being integrated into large-scale malware analysis pipelines to expedite day-to-day analyst work, enhance the speed and quality of related malware queries, and improve results for ML tasks. Embedding AV scan data using AVScan2Vec provides a number of advantages over existing malware feature extraction methods:
\vspace*{-0.1cm}
\begin{enumerate}

\item \textbf{Supports malware in different file formats.} AV products can detect malware which targets a variety of operating systems. AVScan2Vec can embed any malware sample that is detected by one or more AV products. Many other feature extraction approaches are limited to malware in a particular file format (e.g. Windows Portable Executable files). 
\vspace*{0.1cm}

\item \textbf{Fully-systematic feature learning.} AVScan2Vec performs self-supervised feature learning without any manual influence over which features are selected. This avoids continuous work to update feature extractors (because malware is written by an adversarial agent) and instead leverages the work that is already being performed by a multitude of AV companies.
\vspace*{0.1cm}

\item \textbf{Easy-to-obtain data.} 
AVScan2Vec does not require access to raw malicious binaries. AV scan data can be readily obtained online from sources such as VirusTotal \cite{virustotal}. In environments where malware cannot be uploaded externally, collections of local AV products can be used. Larger organizations likely have sizeable collections of AV scan data from one or both sources already, enabling straightforward integration of AVScan2Vec into an analysis pipeline.
\vspace*{0.1cm}

\item \textbf{Low storage and computation overhead.} AV scan data is considerably smaller than raw malicious files and does not incur much storage overhead in comparison. Additionally, AVScan2Vec vectors are smaller than leading vector formats. AVScan2Vec can ingest and vectorize AV scan data using a single GPU.
\vspace*{0.1cm}

\item \textbf{Scales to massive dataset sizes.} AV scan data is an inexpensive and scalable source of high-level malware features. Approaches that require dynamic analysis, disassembly, or other time-consuming types of analysis cannot feasibly scale to production malware corpora sizes.

\end{enumerate}

In the remainder of this section, we will briefly review the limited related work to our own and the nature of AV scan reports, since they may be unfamiliar to those who are not a part of the malware analysis domain. Subsequently, we define our AVScan2Vec  architecture in Section \ref{sec:model-architecture}. This requires implementing a parser to tokenize the unusual, unspecified, and product-dependent labels that are provided in each report. The design of the AVScan2Vec architecture carefully considers the structure of AV scan data and accounts for challenges such as the enormous number of unique tokens in AV labels. To train the model, we use a two-step masked token prediction pre-training phase followed by Multiple Negatives Ranking loss in the fine-tuning phase. In Section \ref{sec:pre-training}  we review the details of pre-training AVScan2Vec on over 30 million reports, and review our validation procedures to ensure that the model is robust and information leakage did not occur. Section \ref{sec:finetune-eval} then fine-tunes the model on 10 million pairs of AV scan reports and performs experimental evaluation. This is done with multiple relevant information retrieval tasks that a malware analyst would regularly perform, and we show that: 1)  AVScan2Vec significantly outperforms other approaches when used to train feed-forward neural networks that classify malware by family and by behavior. 2) AVScan2Vec dominates all other approaches in Homogeneity, Completes, and V-Measure when used by two common algorithms for clustering malware. 3) AVScan2Vec vectors can be quickly computed using a single GPU and work well with metric acceleration structures to enable 16 millisecond search times over 7 million samples. 

\subsection{Related Work}
\label{sec:related-work}

\citet{saxe2015} extracted features from byte entropy statistics, function imports, and Windows Portable Executable (PE) header metadata. \citet{ember}'s feature vector format (EMBER) used the prior features in addition to string statistics and additional file metadata. However, these two feature extraction methods cannot be used for other file formats, or for PE files with missing or corrupt metadata. Furthermore, due to the use of unscaled, tabular data, they are less suitable for deep learning and applications that require pairwise distance computation. Despite this, EMBER has become a leading approach, popularized by its speed and ease of use. 

\citet{bwmd} applied the Burrows-Wheeler transform to raw byte contents to obtain a feature vector. This method (BWMD) does not rely on hand-picked features and is not restricted to any particular file format. Furthermore, BWMD maps files into Euclidean space, a property that facilitates fast and effective nearest neighbor lookup. A major restriction is BWMD's 65536-dimensional vector (85$\times$ larger than AVScan2Vec's). Furthermore, techniques such as packing, obfuscation, and polymorphism can be used to produce files with identical behavior yet very distinct BWMD vectors. 

\citet{islam2013} and \citet{shijo2015} introduced feature extraction techniques that integrate features from both static and dynamic analysis. \citet{ahmadi2016} performed feature extraction for PE files using byte n-grams, file metadata, and disassembly. \citet{yuxin2019} disassembled and constructed control flow graphs for unpacked PE files, then extracted opcode n-gram sequences. They trained a deep belief network to perform feature learning on the opcode n-grams. We do not consider feature extraction methods that rely on disassembly or dynamic analysis because featurizing all malicious files in a  production-scale dataset is infeasible when using them.

\subsection{Antivirus Scan Data}

The remainder of Section \ref{sec:introduction} will provide additional details about AV scan data and introduce important concepts.

\subsubsection{Antivirus labels}
When an AV product detects a malicious flag, it outputs an \textbf{\textsl{AV label}}. The label is a sequence of tokens, each of which describes an attribute of the file. These attributes may include the operating system or architecture the malware targets, behaviors the malware may exhibit, the family or broad category to which the malware belongs, a threat group or campaign to which the malware is attributed to, a vulnerability or exploit associated with the malware, the programming language in which the malware was implemented, or the packer used to pack the malware \cite{avclass2}.

An example AV label is \textsl{Ransom.Win32.Wanna.xyz!gen}. The ``Ransom" token indicates that the malware belongs to the ransomware category of malware. The ``Win32" token indicates that the malware targets the Windows 32-bit platform. The ``Wanna" token indicates the malware family that the file belongs to (Wanna is an alias for the WannaCry family). The ``xyz" suffix indicates the specific means of detection (i.e. the AV product may employ multiple signatures, heuristics, or other technologies for detecting WannaCry) \cite{hahn}.
Finally, ``gen" is a modifier token which indicates that a generic detection method was used, such as a heuristic.

\subsubsection{Antivirus Scan Reports}
Using labels from a single 

\begin{wrapfigure}[16]{R}{0.5\textwidth}    
    \vspace*{-26pt}
    \centering
\includegraphics[width=0.48\columnwidth,keepaspectratio]{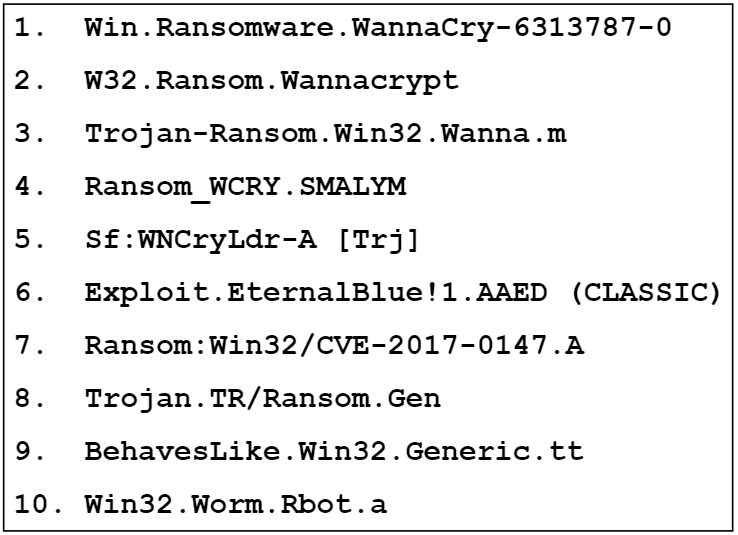}
    \caption{Fictitious AV scan report for a file detected as malware by 10 AV products. Labels 1-5 classify this file into the WannaCry family. Labels 6-7 convey that the malware leverages EternalBlue in order to exploit the vulnerability CVE-2017-0147.
    Labels 8-9 state that the file is ransomware targeting 32-bit versions of Windows. Label 10 incorrectly classifies the file into the Rbot family.}
    \label{fig:av-results}
\end{wrapfigure}

\noindent AV product is widely known to produce results that may be noisy, inconsistent, or ambiguous. Instead, scanning malware with a collection of AV products provides a more comprehensive view of its attributes. We refer to the collective results from multiple AV products for a single malware sample as an \textbf{\textsl{AV scan report}}. This is the format which modern AV-based tagging tools \cite{avclass, avclass2}, as well as AVScan2Vec, use as input. An example AV scan report is displayed in Figure \ref{fig:av-results}. Note that each AV product uses its own label format and naming conventions. Interpreting the contents of an AV scan report may be difficult due to inconsistent naming, differing granularity, and potentially erroneous information. Each label in an AV scan report originates from a particular signature, heuristic, or other type of detection technology employed by an AV product. If the same AV label is present in two AV scan reports, then the corresponding malware samples must share some common latent attribute(s) identified by an AV product. If two AV scan reports have many similar or identical AV labels, then the corresponding malware samples likely share many common latent attributes. This idea is the foundation of our work:

\vspace*{0.1cm}

\begin{description}[leftmargin=0pt]
    \item \textbf{\textsl{AV labels encode latent information about their corresponding files, and two malware samples with similar AV scan reports are themselves intrinsically similar.}}
\end{description}

\section{AVScan2Vec Model Architecture}
\label{sec:model-architecture}

AVScan2Vec is a language model tasked with interpreting the (often heterogeneous) AV labels contained within AV scan reports.
This section discusses the AVScan2Vec architecture, which is logically divided into data preprocessing, pre-training, and fine-tuning phases.
Data preprocessing is necessary for converting the contents of an AV scan report into a sequence of token embeddings. Pre-training enables AVScan2Vec to learn semantic meaning for individual tokens. 
The fine-tuning phase teaches AVScan2Vec to embed entire AV scan reports into individual vectors, where related malware samples are embedded into nearby vectors.

\subsection{Data Preprocessing}
\label{sec:preprocessing}

A large dataset of AV scan reports must be preprocessed in order to train AVScan2Vec.
The design choices for AVScan2Vec's preprocessing stage and model architecture enable it to handle the long sequences and very large vocabulary sizes inherent to this learning task. AV scan reports may require sequence lengths of several hundred tokens, and the dataset used to pre-train AVScan2Vec contains more than 40 million distinct tokens - over 1000$\times$ the vocabulary size of BERT \cite{bert}. Other considerations, such as how to represent AVs that did not participate in a scan or that detected a file as benign, were also incorporated into AVScan2Vec's preprocessing phase.

\subsubsection{Antivirus Label Preprocessing}

\begin{wrapfigure}[23]{R}{0.5\textwidth}    
    \vspace*{-12pt}
    \centering
    \includegraphics[width=0.47\columnwidth,keepaspectratio]{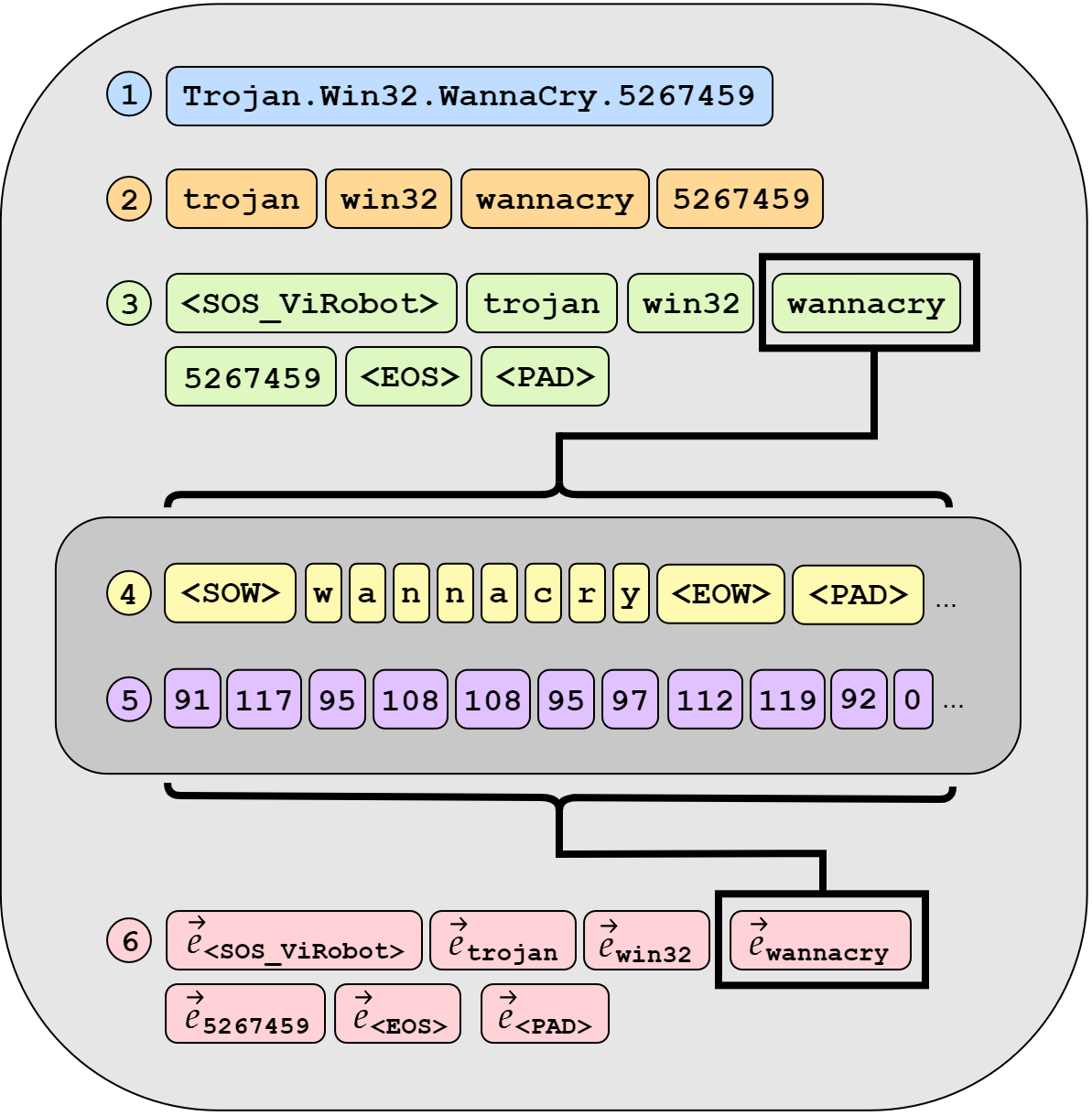}
    \caption{Preprocessing of an AV label produced by the ViRobot AV product. (1) shows the original label. (2) shows the label after it has been tokenized and normalized. (3) displays the addition of special tokens indicating the start of the label, end of the label, and padding. (4) shows how the wannacry token is separated into individual characters, with the addition of start of word, end of word, and padding tokens. (5) shows the numeric representation of each character and special token. (6) displays the CharacterBERT-style embeddings of each token.}
    \label{fig:tokenizing}
\end{wrapfigure}

Figure \ref{fig:tokenizing} shows an example AV label being preprocessed. The original AV label (1) is split on non-alphanumeric characters and each token is converted to lowercase (2). Then, the label is truncated to five tokens. This is necessary for reducing memory usage due to long sequence lengths. We observe that 95.17\% of AV labels in our dataset are five or fewer tokens, and tokens which indicate the behavior, target architecture, and family are almost always placed near the beginning of the label. Therefore, we judge this to be an acceptable design compromise. 

Labels may be missing from a scan report due to an AV abstaining or detecting the file as benign. The \textless ABS\textgreater\space and \textless BEN\textgreater\space special tokens are used in each of these cases, respectively. Each AV product is assigned a special \textless SOS\textgreater\space token which indicates the start of the label it produced. For example,  \textless SOS\_ViRobot\textgreater\space indicates the start of the ViRobot AV product's label in the sequence. The \textless SOS\textgreater\space and \textless EOS\textgreater\space tokens are placed before and after the label's tokens, respectively, and then \textless PAD\textgreater\space tokens are used to extend the label to its maximum length (3). The maximum length is a hyperparameter named $L$, which defaults to seven (up to five tokens from the AV label plus \textless SOS\textgreater\space and \textless EOS\textgreater).

\subsubsection{Token Embedding}
\label{sec:token-embedding}

After each AV label is converted to normalized tokens, it must be embedded to be used as input to AVScan2Vec. However, the extreme token vocabulary size of this learning task makes a traditional embedding approach infeasible. An additional concern is that new AV scan data may contain tokens outside of a pre-established vocabulary. To circumvent both of these issues, AVScan2Vec employs the embedding method introduced by CharacterBERT \cite{characterbert}, which supports a large, open-set vocabulary. As shown in Figure \ref{fig:tokenizing}, each token is split into individual characters (4). Tokens that are longer than 20 characters are truncated, and \textless SOW\textgreater\space and \textless EOW\textgreater\space tokens are placed on either side to indicate the start and end of the token. Each lowercase letter, numeral, and special token is mapped to a numeric representation (5). This sequence is used as input to a series of 1-D Convolutional Neural Networks (CNNs) with different filter sizes, followed by two Highway layers and a projection into a single vector \cite{characterbert, highway}. We use $\vec{e}$ to denote token embeddings (6).

AVScan2Vec's token embedding implementation is identical to the original CharacterBERT implementation, with the exception of minor hyperparameter changes. Due to nearly all (99.98\%) of tokens being shorter than 20 characters, AVScan2Vec uses a maximum character length of 20 rather than 50, and the number and size of of CNN filters was adjusted accordingly. Position and segment embeddings are added to the token embeddings to provide information about token ordering and which AV product produced each token \citep{vaswani}.

\subsection{Pre-training Tasks}

\begin{wrapfigure}[19]{R}{0.5\textwidth}    
    \vspace*{-12pt}
    \centering
    \includegraphics[width=0.47\columnwidth,keepaspectratio]{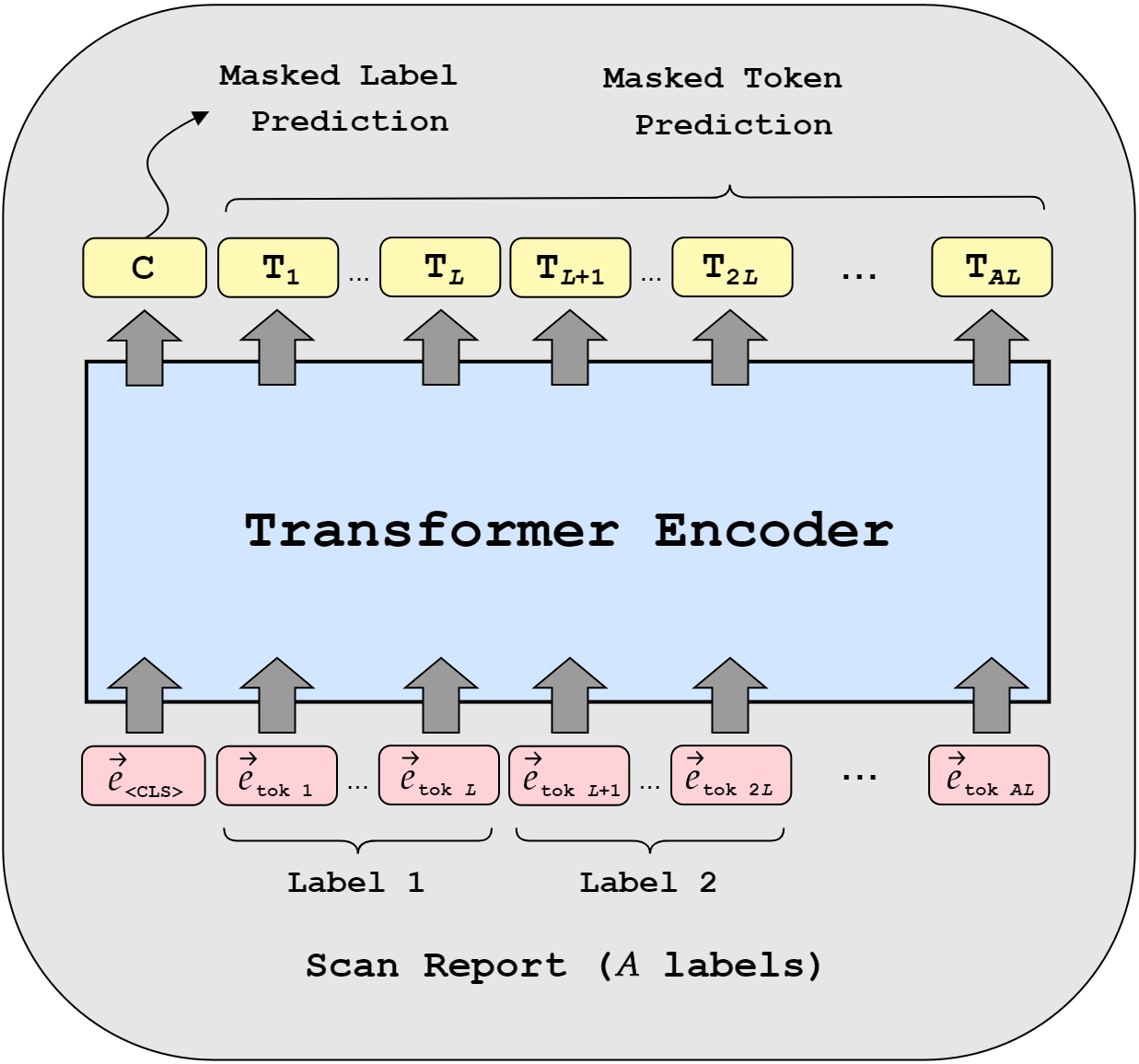}
    \caption{AVScan2Vec pre-training pipeline. Each preprocessed scan report consists of $\vec{e}_{<CLS>}$ followed by the embedded tokens $\vec{e}_{tok 1} \; ... \; \vec{e}_{tok AL}$. A Transformer encodes these tokens. The final hidden states corresponding to $\vec{e}_{<CLS>}$ are denoted $C$ and are used for Masked Label Prediction. The hidden states for $\vec{e}_{tok 1} \; ... \; \vec{e}_{tok AL}$ are denoted $T_{1} ... T_{AL}$ and are used for Masked Token Prediction.}
    \label{fig:pretrain}
\end{wrapfigure}

Figure \ref{fig:pretrain} shows how AVScan2Vec is pre-trained to learn token semantics from AV scan data. Each input sequence to AVScan2Vec represents a single AV scan report. The first token in each input sequence is the embedding of a special \textless CLS\textgreater\space token, indicating the start of the sequence and denoted $\vec{e}_{<CLS>}$. The remainder of the input sequence consists of token embeddings $\vec{e}_{tok 1} \; ... \; \vec{e}_{tok AL}$, where $A$ is the number of AV products. Each AV product's label contributes $L$ token embeddings (as described in Section \ref{sec:token-embedding}) and always appears at a fixed location within every sequence. This results a total sequence length of $A \cdot L$ + 1.

A Transformer encoder is trained using batches of these sequences as inputs. AVScan2Vec's default hyperparameters employ four Transformer block layers, hidden size $D$ = 768, and eight attention heads per layer \cite{vaswani}. The final hidden state corresponding to $\vec{e}_{<CLS>}$ is denoted $C$ and is used for the Masked Label Prediction pre-training task. The remainder of the final hidden states are denoted $T_{1} ... T_{AL}$, and are used for Masked Token Prediction.

\subsubsection{Masked Token Prediction}

Masked Token Prediction teaches AVScan2Vec the semantic meanings of tokens by making inferences based on AV scan report contents. During this task, approximately 5\% of tokens in each sequence are selected at random and held out for prediction. AVScan2Vec must use the surrounding tokens - from both the current label and the remainder of the sequence - as context for making informed predictions. Like BERT, the selected token has an 80\% chance of being replaced with a special \textless MASK\textgreater\space token, a 10\% chance of being replaced with a random token, and a 10\% chance of no modification \cite{bert}. To prevent AVScan2Vec from ``cheating," any other tokens in the sequence which are identical to the chosen token are also replaced with \textless MASK\textgreater. This encourages the model to learn context from tokens that have related meanings, such as family aliases.

If the $i^{th}$ token in the sequence is selected, then the final hidden state $T_{i}$ is used as input to a small feed-forward neural network (FFNN), followed by adaptive softmax approximation to obtain log probabilities for the 10 million most common tokens \cite{alswl}. This softmax approximation strategy is necessary due to the extreme number of distinct tokens in AV scan data. Negative log-likelihood loss is computed using these log probabilities and the selected token.

\subsubsection{Masked Label Prediction}

Masked Label Prediction teaches AVScan2Vec semantic meanings for entire AV labels. During this task, one AV product that scanned the file is selected at random and its tokens are replaced with \textless ABS\textgreater\space special tokens. It is possible for an AV product that detected the file as benign to be selected. 
A LSTM decoder is trained to autoregressively predict the tokens in the held-out AV label. The decoder has a hidden size of $D$
(768 by default) and $n\_layers=4$ recurrent layers. The initial input to the LSTM decoder is the embedding of the \textless SOS\textgreater\space token for the AV whose label must be predicted (e.g. $\vec{e}_{<SOS\_ViRobot>}$). Recall that $C$ is the final hidden state of the Transformer obtained from $\vec{e}_{<CLS>}$. $C$ is used as input to small FFNN with an input size of $D$ and an output size of $D \cdot n\_layers$. The output of this FFNN is reshaped and used as the initial hidden state of the LSTM before the first timestep. The initial cell state of the LSTM is set to zero before the first timestep.

At each decoding timestep $t$, the LSTM's outputs are passed to another small FFNN followed by adaptive softmax approximation, resulting in log probabilities for the 10 million most common tokens. The resultant hidden and cell states $h_{t}$ and $c_{t}$ also update at each decoding timestep, and they are used as the initial hidden and cell states of timestep $t+1$. Iteration continues until the token prediction network produces an \textless EOS\textgreater\space token or until $L$ timesteps pass. The decoder uses 50\% teacher forcing to assist with training. In the 50\% of cases where teacher forcing is not used, the token with the highest log probability is used as input to the LSTM decoder during timestep $t+1$. We observed that the model achieves highest performance by always selecting the token with the highest likelihood, so strategies such as beam search were not necessary.

\subsection{Fine-tuning Tasks}
\label{sec:fine-tuning}

\begin{wrapfigure}[17]{R}{0.5\textwidth}    
    \vspace*{-12pt}
    \centering
    \includegraphics[width=0.47\columnwidth,keepaspectratio]{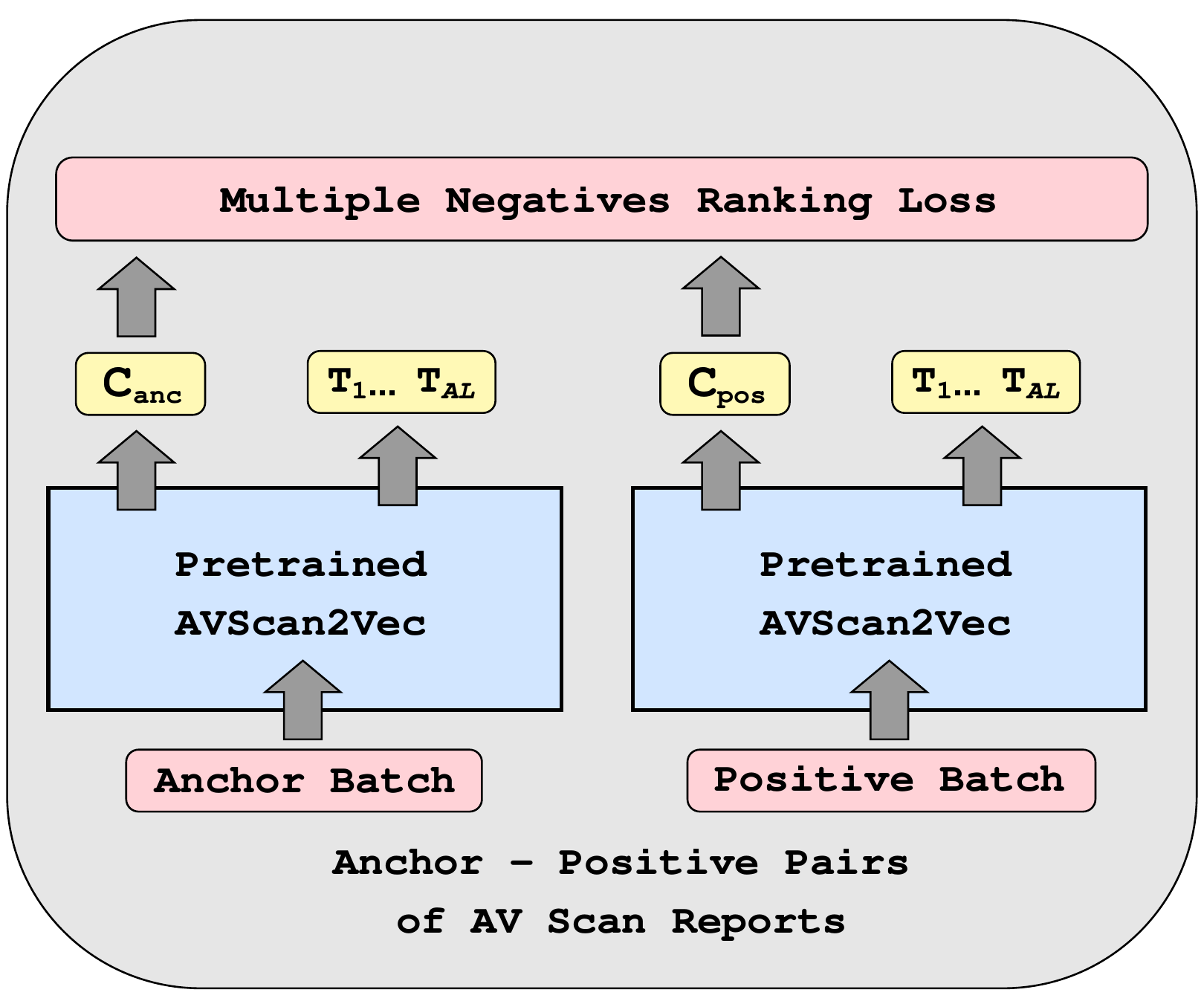}
    \caption{AVScan2Vec fine-tuning pipeline. Batches of anchor-positive pairs are input to the pre-trained AVScan2Vec Siamese network. $\textbf{C}_{\text{anc}}$ and $\textbf{C}_{\text{pos}}$ are used to compute MNR loss for the batch, while the remainder of the final hidden states are ignored.}
    \label{fig:finetune}
\end{wrapfigure}

Once AVScan2Vec has been pre-trained to learn the general semantics of tokens within AV labels, it can be fine-tuned for high performance on other tasks. In our case, we fine-tune AVScan2Vec to learn a vector representation for an entire scan report. A Siamese network structure (e.g. two pre-trained AVScan2Vec models with shared weights) is used during fine-tuning. Batches of $k$ pairs of scan reports are provided as input. The first scan report in each pair is called the anchor, and is randomly sampled from the fine-tuning dataset. The second scan report in each pair is the positive, which is the scan report for another malware sample in the same family as the anchor. We describe our procedure for identifying anchor-positive pairs in Section \ref{sec:finetune-data}.

The AVScan2Vec fine-tuning objective is to minimize the Multiple Negatives Ranking (MNR) loss \cite{mnrloss}. Let $\textbf{C}_{\text{anc}} = (C_{\text{anc}_1}, ..., C_{\text{anc}_k})$ be the final hidden states obtained from the $\vec{e}_{<CLS>}$ token for each anchor scan report in the batch, and let $\textbf{C}_{\text{pos}} = (C_{\text{pos}_1}, ..., C_{\text{pos}_k})$ be the same for each positive scan report in the batch.  For each anchor $C_{\text{anc}_i} \in \textbf{C}_{anc}$, every $C_{\text{pos}_j} \in \textbf{C}_{pos}$ in the batch is treated as a negative candidate (i.e. unrelated to the anchor) if $i \neq j$. The model learns to reduce the distance between each anchor-positive pair, while increasing the distance between each anchor and its negative candidates. Define a score function $S\left(C_{\text{anc}_i}, C_{\text{pos}_j}\right)$ as the dot product of $C_{\text{anc}_i}$ and $C_{\text{pos}_j}$ and let $\theta$ be the model parameters. Formally, the MNR loss for a batch is given by \cite{mnrloss}:

\begin{equation*}
    \mathcal{J}_\theta(\textbf{C}_{\text{anc}}, \textbf{C}_{\text{pos}}) =-\frac{1}{k} \sum_{i=1}^{k} \left[ S\left(C_{\text{anc}_i}, C_{\text{pos}_i}\right)-\log \sum_{j=1}^{k} e^{S\left(C_{\text{anc}_i}, C_{\text{pos}_j}\right)} \right]
\end{equation*}\

\vspace*{8pt}

MNR loss is especially effective for malware-related tasks due to the field's severe lack of labeled data. This is because determining if two malware samples are related (i.e. an anchor-positive pair) is much less difficult than determining if they are unrelated (i.e. an anchor-negative pair). The former task can be achieved using robust file similarity metrics, while the latter requires family labels. MNR loss does not require negative candidates to be explicitly provided and instead takes advantage of the considerable diversity of the problem space. Due to the vast number of malware families in existence, the negative candidates in a batch are very unlikely to belong to the same family as the anchor.

\section{Pre-Training AVScan2Vec}
\label{sec:pre-training}

AVScan2Vec was pre-trained on more than 33.4 million AV scan reports. This enormous dataset size is fundamental in AVScan2Vec's capacity to comprehend AV label semantics and generalize to new data. This section provides technical details regarding how AVScan2Vec was pre-trained. It also describes our validation process.

\subsection{Pre-Training Implementation}

\subsubsection{Data collection.}
Training AVScan2Vec requires a large dataset of AV scan reports. To collect this data, we used VirusTotal, an online malware analysis platform \cite{virustotal}. Although scan data for training AVScan2Vec could have been obtained by manually scanning malware with AV products, we elected to use VirusTotal because an enormous amount of AV scan data was readily available through it. We queried the VirusTotal API for scan reports for 40,307,433 malware samples in chunks 0 through 465 of the VirusShare dataset \citep{virusshare}. 
Scan reports from VirusTotal contain the latest AV results available for a malware sample, but it may have been months or even years since a malware sample's last scan. Dates of the scan reports we collected range from May 2006 to Apr. 2023. The set of AV products included in VirusTotal has gradually changed over time and we identified 89 distinct AVs that were present in at least 1\% of the queried scan reports. A list of these AVs has been provided in Table \ref{tab:avscan2vec_av} in Appendix A. In order to train AVScan2Vec, it was necessary to discard any scan reports that did not include at least two detections from the 89 supported AVs, or that were missing other necessary information. This resulted in a pre-training dataset of 37,132,493 total scan reports.

\subsubsection{Technical Details}

AVScan2Vec was pre-trained using eight NVIDIA Quadro RTX 8000 GPUs. 33,419,200 ($\approx$90\%) scan reports  from the VirusShare dataset were used as a training set and 3,713,200 ($\approx$10\%) were used for validation. Batches of 100 scan reports were provided to each GPU in parallel, and gradients obtained from the eight GPUs were averaged for the backward pass. The AdamW optimizer was used with an initial learning rate of 2.5$e^{-4}$, a cosine annealing learning rate scheduler, and a warm learning rate restart at the beginning of every epoch \cite{cosineannealing}. The model converged after training for 12 epochs (approximately 12 days). The pre-trained AVScan2Vec model achieved an overall Masked Token Prediction accuracy of 91.02\% and Masked Label Prediction accuracy of 83.97\%. AVScan2Vec's lowered

\begin{wraptable}[7]{R}{0.5\textwidth}
\vspace{-10pt}

\centering
\resizebox{0.375\columnwidth}{!}{%
\begin{tabular}{@{}lr@{}}
\toprule
Pre-Training Task & Accuracy\\
\midrule
    Masked Token Prediction & 91.02\%\\
    Masked Label Prediction & 83.97\%\\
\bottomrule
\end{tabular}
}
\label{tab:overall_accuracy}
\vspace*{6pt}
\caption{AVScan2Vec Pre-Training Accuracy}

\end{wraptable}

\noindent 
accuracy for Masked Label Prediction is explained by the increased difficulty of the task. Regardless, these results show that AVScan2Vec 
undeniably has an understanding of the semantic meaning of tokens in AV scan data, the label formats for each AV, and how to make inferences based on scan report contents. 

\subsection{Pre-training Validation}
\label{sec:pretrain-eval}
To ensure that AVScan2Vec is suitable for later fine-tuning tasks under a wide range of circumstances, we performed multiple validation experiments. The main objective of our validation was to mitigate the risk of unexpected or irregular behavior when AVScan2Vec receives sparse or out-of-distribution (OOD) inputs. We analyzed multiple scenarios which can cause AVScan2Vec's pre-training tasks to become more difficult, with the expectation that AVScan2Vec's predictive ability does not "break", even if accuracy degrades somewhat.

\subsubsection{Accuracy Per Antivirus Product}

Each AV product uses its own label format and naming conventions. Furthermore, some AV products rarely occur in the pre-training dataset (often due to being included in VirusTotal for a limited period of time). It is important to ensure that there are no AV products for which AVScan2Vec cannot learn the pre-training tasks. To confirm this, we measured AVScan2Vec's accuracy when predicting tokens and labels for each AV product using the $\approx$ 3.7 million scan reports in the test set.

Masked Token Prediction accuracy ranged from 79.64\% (Baidu International) to 100.00\% (Cylance) and Masked Label Prediction accuracy ranged from 66.91\% (Norman) to 99.35\% (Babable). The full results are listed in Table \ref{tab:avscan2vec_av} in Appendix A. We observed that accuracy is highest for AV products with simple label formats (e.g. APEX, Babable, Cylance, Palo Alto, and SentinelOne). Groups of AV products known to make correlated labeling decisions also scored highly (e.g. Ad-Aware, Arcabit, Emsisoft, Fireye, F-Secure, GData, and MicroWorld-eScan use the BitDefender engine) \cite{zhu2020}. As expected, accuracy degrades for some AV products with difficult label formats and/or a lack of data. However, given that AVScan2Vec can select from over 10 million tokens during inference, it is evident that AVScan2Vec learned the pre-training tasks for all 89 AV products. 
We would like to emphasize that AVScan2Vec's Masked Token Prediction and Masked Label Prediction accuracies are not related to an AV product's performance at detecting and labeling malware.

\vspace*{6pt}
\subsubsection{Accuracy Per Number of Positive Detections in Scan Report}

\begin{wrapfigure}[16]{R}{0.5\textwidth}    
    \vspace*{-24pt}
    \hspace*{-8pt} \includegraphics[width=0.55\columnwidth,keepaspectratio]{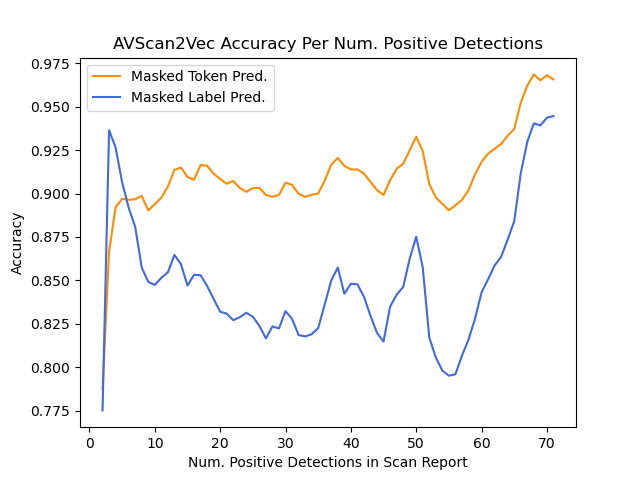}
    \caption{Masked Token Prediction and Masked Token Prediction accuracy of AVScan2Vec per number of AV products in the scan report which detect the file as malware. AVScan2Vec is still able to perform both tasks even when it can only observe a single AV label.
    }
    \label{fig:avscan2vec_detect}
\end{wrapfigure}

Most scan reports from VirusTotal contain scan results from approximately 70 AV products. However,
only the AV products with positive detections (i.e. they detected the file as malware) can produce AV labels for a scan report. We hypothesized that AVScan2Vec's performance may decrease for scan reports where very few AVs detect the file as malware, due to limited information. To test this, we computed accuracy for both pre-training tasks with respect to the number of positive detections in the scan report. Results are shown in Figure \ref{fig:avscan2vec_detect}. In the worst case (when the scan report only contains two positive detections), AVScan2Vec achieves above 78\% accuracy for Masked Token Prediction and above 77.5\% accuracy for Masked Label Prediction. AVScan2Vec is only able to observe a single AV label, as the other is held out for Masked Label Prediction. We conclude that AVScan2Vec is still able to effectively perform both pre-training tasks even under restrictive conditions when the minimum amount of data is available in a scan report.

\vspace*{4pt}
\subsubsection{Accuracy Per Month of Scan Report}

\begin{wrapfigure}[14]{R}{0.5\textwidth}    
    \vspace*{-40pt}
    \hspace*{-8pt}
    \includegraphics[width=0.55\columnwidth,keepaspectratio]{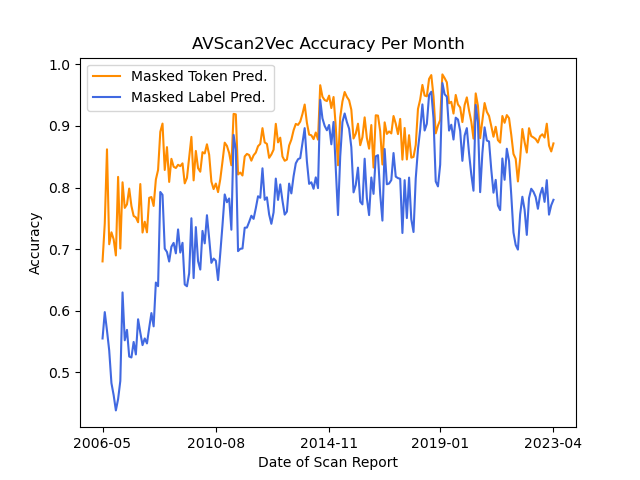}
    \caption{Accuracy of AVScan2Vec per month the scan report was created. Accuracy generally stays high, although it degrades in months where there is limited training data.}
    \label{fig:avscan2vec_date}
\end{wrapfigure}

The pre-training dataset includes scan reports from nearly a 17 year span (May 2006 to Apr. 2023). The malware ecosystem changes constantly and quickly, with new families  being introduced and old families falling out of favor \cite{r1sm}. Additionally, there are months where limited training data was available. It is important to understand how both of these factors impact AVScan2Vec. AVScan2Vec's monthly accuracy on both pre-training tasks is displayed in Figure \ref{fig:avscan2vec_date}. AVScan2Vec's low accuracy before 2009 likely results from insufficient amounts of training data, as some AV products, malware families, and labels only appear in the training set prior to 2009 

\newpage
\noindent and the training set only included approximately 5,000 -

\noindent 10,000 scan reports per month during this interval. However, since AVScan2Vec can select from over 10 million tokens during inference, we consider these accuracies indicative of AVScan2Vec's continued ability to perform both pre-training tasks under limited training data (albeit with lowered accuracy, which is understandable).

\subsubsection{Simulating Performance on ``Future" Malware}
To model AVScan2Vec's accuracy when encountering scan reports for novel malware in a real-world setting, we pre-trained the model a second time using a temporal split of the VirusShare dataset. The training set was composed of 33,419,200 scan reports from May 2006 to June 2021, while the test set included 3,713,200  scan reports from June 2021 to Apr. 2023. The rapidly-changing nature of the malware ecosystem results in new malware families

\begin{wrapfigure}[13]{R}{0.5\textwidth}    
    \vspace*{-38pt}
    \hspace*{-8pt} 
    \includegraphics[width=0.55\columnwidth,keepaspectratio]{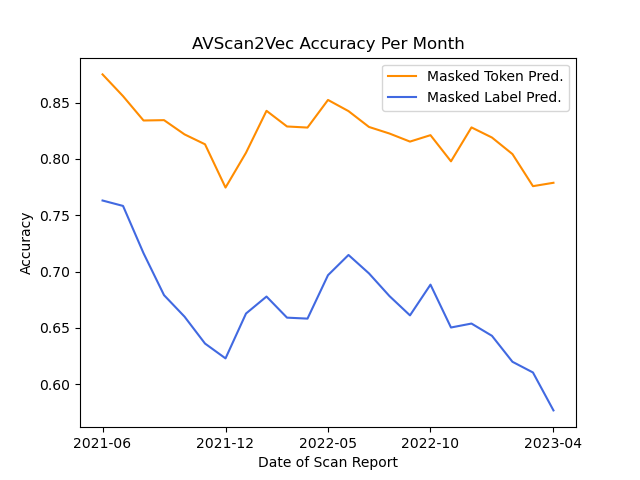}
    \caption{Accuracy of AVScan2Vec trained on a temporal split per month the scan report was created. AVScan2Vec is still successful at both pre-training tasks for data nearly two years into the ``future".}
    \label{fig:avscan2vec_date_temporal}
\end{wrapfigure}

\noindent and new antivirus labels, even within the span of a single month \cite{jordaney2017transcend, r1sm}. Therefore, the test set includes malware which did not yet exist during the time period of the training set, and the proportion of novel malware increases with the date. This scenario allows us to simulate the performance of AVScan2Vec on OOD data. Figure \ref{fig:avscan2vec_date_temporal} reflects the expected decline in performance as AVScan2Vec encounters more OOD data. However, given that AVScan2Vec can choose from over 10 million tokens at inference time, its predictive abilities clearly remain robust - even when making predictions on scan data nearly two years into the "future". For best results, we recommend that AVScan2Vec should be regularly re-trained with updated scan reports.

\section{Fine-tuning AVScan2Vec}
\label{sec:finetune-eval}

AVScan2Vec was fine-tuned using 10 million anchor-positive pairs of AV scan reports. This section includes technical details about how AVScan2Vec was fine-tuned, followed by evaluation results that compare the AVScan2Vec vector to other malware vector representations.

\subsection{Fine-Tuning Implementation}

\subsubsection{Anchor-Positive Selection}
\label{sec:finetune-data}

The main objective of AVScan2Vec is to embed the scan reports of similar malware samples into nearby vectors. However, even small alterations to a malicious file may alter the label that an AV assigns to it. Furthermore, we wish for AVScan2Vec to be resilient against external factors such as the version of the AV product and the set of AV products included in the scan report. In order to remain resilient against these perturbations, it is necessary to fine-tune AVScan2Vec after pre-training has completed.

In order to fine-tune AVScan2Vec, anchor-positive pairs of malware samples must first be identified. To do so, we employed the Trend Locality Sensitive Hash (TLSH), which can identify files with similar contents \cite{tlsh}. We considered two files with a TLSH distance score less than 30 to be related. \citet{tlsh} evaluated this distance threshold to have a false positive rate of just 0.00181\%. It is possible that large edits, packing, polymorphism, or other obfuscation may cause related files to not be identified by TLSH. However, we judge that our experimental setup is most effective when false positives are minimized and that some false negatives are allowable. Our approach still enables a large set of related files to be identified with very high fidelity, which is sufficient for fine-tuning. We identified 26,274,859 pairs of similar files using this method, taking care not to permit duplicate pairs. We then randomly downsampled to 10,010,000 pairs.

\subsubsection{Technical Details}
AVScan2Vec was fine-tuned using a single NVIDIA Quadro RTX 8000 GPU. 10,000,000 anchor-positive pairs of scan reports were used as a training set, with an additional 10,000 for validation purposes. The model used a batch size of 100, the AdamW optimizer, and an initial learning rate of 1$e^{-6}$. The learning rate was decreased when validation loss plateaued. The model fine-tuned for one epoch (approximately 37 hours), after which it began to overfit and training was halted.

\subsection{Fine-Tuning Evaluation}

Next, we evaluated the performance of AVScan2Vec vectors on relevant ML tasks and compared its performance against BWMD and EMBER, which are leading malware feature extraction approaches. BWMD excels when used for malware nearest-neighbor lookup and is not limited by file format. The EMBER vector format has been widely adopted as a means of featurizing malicious PE files. Because EMBER vectors are not normalized by default and contain categorical data, we normalized and applied PCA to each vector, preserving 90\% of explained variance. We repeated each experiment using both the default EMBER vectors and the normalized vectors (labeled as EMBER-N). Because scalability is a focal point of our work, we do not consider feature extraction methods which cannot scale to production-size malware corpus sizes, including approaches that rely on dynamic analysis and disassembly. Giving two-minute sandbox runs to the $\approx$37 million files in our pre-training dataset would require $\approx$141 compute-years, and use of tools such as IDA Pro for disassembly would be comparatively expensive. We evaluated AVScan2Vec using three common ML tasks that are frequently performed on malware data: nearest-neighbor lookup, classification, and clustering. We tested multiple common algorithms for each task and ensured that our evaluations used publicly-available benchmarks when possible.

\subsubsection{Evaluation Datasets}

We selected the MOTIF \cite{motif} and SOREL \cite{sorel} benchmark datasets for evaluating AVScan2Vec after it was fine-tuned. We took care not to use evaluation datasets that were labeled using AV-derived data, such as from tools like AVClass \cite{avclass} or AVClass2 \cite{avclass2}. This would have been a source of bias given that AVScan2Vec is trained on AV-derived data as well.

MOTIF is the largest public dataset of malware with full ground-truth family labels. Each malware sample in MOTIF was labeled using manual analysis \cite{motif}. In the experiments which used MOTIF, malware samples with invalid VirusTotal scan reports and whose families only occurred once in the dataset were discarded, resulting in 2,962 malware samples from 323 families.

The SOREL dataset includes 9,919,065 malicious PE files available for download. SOREL is labeled according to 11 separate malicious behaviors. A malware sample may have multiple behavioral tags, as malware frequently displays multiple types of malicious behaviors. 
SOREL was labeled according to multiple static rules in combination with information internal to Sophos-RevesingLabs \cite{sorel}. Not all of the malware in the SOREL dataset is available on VirusTotal. We were able to obtain AV scan reports for 7,292,023 of the 9,919,065 files. Some of the following experiments use all 7,292,023 of these files (which we refer to as the "full" SOREL dataset). Since there is a large class imbalance within SOREL, some experiments use a randomly down-sampled dataset with 110,000 files from SOREL, ensuring that each tag occurs at least 10,000 times (which we call SOREL-110000).

\begin{wraptable}[12]{R}{0.5\textwidth}
\vspace{-12pt}
\centering

\resizebox{0.28\columnwidth}{!}{%

\begin{tabular}{@{}lr@{}}
\toprule
Representation & Size \\ \midrule
Raw Binaries & 13.9 TB \\
\textbf{VirusTotal Reports} & \textbf{140.6 GB}     \\
BWMD Vectors & 1.9 TB \\
EMBER Vectors & 69.4 GB              \\ 
\textbf{AVScan2Vec Vectors}         & \textbf{22.4 GB}      \\ \bottomrule
\end{tabular}
}
\vspace*{6pt}
 \caption{Sizes of 7,292,023 files from the SOREL dataset in various representations. AVScan2Vec's vector is 600$\times$ more compact than the average file in SOREL, 3$\times$ smaller than EMBER vectors, and 85$\times$ smaller than BWMD vectors. These storage savings facilitate faster vector comparisons and lower memory usage for all ML tasks.
}
\label{tbl:sizes}
\end{wraptable}

\subsubsection{Malware Representation Sizes}
\label{sec:representation-size}

Using AVScan2Vec, we vectorized the 7,292,023 VirusTotal reports for SOREL. This took six hours and 16 minutes using a single NVIDIA Quadro RTX 6000 GPU. We estimate that computing BWMD and EMBER vectors for these files would have taken approximately 153 and 161 hours respectively using 72 cores on an Intel Xeon Gold 5520 CPU. This is likely due to a file I/O bottleneck, since raw binaries are much larger than AV scan data (see Table \ref{tbl:sizes}).
BWMD and EMBER vector sizes in Table \ref{tbl:sizes} are extrapolated from SOREL-110000. 
\vspace*{4pt}

\vspace*{12pt}
\noindent
\resizebox{\columnwidth}{!}{%
\begin{minipage}[c]{0.48\textwidth}
\centering
\resizebox{\columnwidth}{!}{%
\begin{tabular}{@{}lrrrr@{}}
\toprule
 & AVScan2Vec & BWMD & EMBER & EMBER-N \\
\midrule
Adware & \textbf{.987} & .971 & .979 & .982 \\
Flooder & \textbf{.987} & .974 & .981 & .986 \\
Ransomware & \textbf{.990} & .975 & .981 & .986 \\
Dropper & \textbf{.973} & .942 & .956 & .959 \\
Spyware & \textbf{.983} & .962 & .969 & .971 \\
Packed & \textbf{.988} & .964 & .978 & .980 \\
Crypto Miner & \textbf{.988} & .974 & .980 & .986 \\
File Infector & \textbf{.990} & .980 & .985 & .987 \\
Installer & \textbf{.988} & .970 & .978 & .984 \\
Worm & \textbf{.987} & .973 & .977 & .980  \\
Downloader & \textbf{.983} & .964 & .972 & .975 \\
\bottomrule
\end{tabular}
}
\vspace*{6pt}
\captionof{table}{FFNN classifier trained for behavioral tag classification. AVScan2Vec has the highest five-fold classification mean ROC-AUC score for all 11 behavioral tags.}
\label{tab:tag_ffnn_accuracy}
\end{minipage}

\hspace*{32pt}

\begin{minipage}[c]{0.48\textwidth}
\centering
\resizebox{\columnwidth}{!}{%
\begin{tabular}{@{}lrrrr@{}}

\toprule
 & AVScan2Vec & BWMD & EMBER & EMBER-N \\
\midrule
Adware & \textbf{.990} & .982 & .988 & .988 \\
Flooder & \textbf{.990} & .984 & .988 & .988 \\
Ransomware & \textbf{.992} & .986 & 990 & .990 \\
Dropper & \textbf{.979} & .960 & .924 & .971 \\
Spyware & \textbf{.986} & .975 & .982 & .981 \\
Packed & \textbf{.990} & .979 & .987 & .987 \\
Crypto Miner & .991 & .984 & .989 & \textbf{.992} \\
File Infector & .993 & .987 & \textbf{.993} & .992 \\
Installer & \textbf{.989} & .984 & .988 & .988 \\
Worm & \textbf{.990} & .982 & .977 & .986  \\
Downloader & \textbf{.988} & .978 & .987 & .983 \\
\bottomrule
\end{tabular}
}
\vspace*{6pt}
\captionof{table}{LightGBM OvR classifier trained for behavioral tag classification. AVScan2Vec has the highest five-fold classification mean ROC-AUC score for nine of 11 behavioral tags.}
\label{tab:tag_lgbm_accuracy}
\end{minipage}
}

\subsubsection{Behavioral Tag Classification Results}
\label{sec:sorel-110000}

Our first experiment tests each vector format's aptitude for representing features related to runtime behavior. We trained FFNN and LightGBM classifiers on SOREL-110000 to predict the 11 SOREL behavioral tags using AVScan2Vec, BWMD, EMBER, and EMBER-N vectors as input. Our FFNN used the same architecture and layer sizes as the FFNN benchmark provided with the SOREL datset \cite{sorel}. Because LightGBM does not support multilabel classification, we trained separate One-vs-Rest (OvR) LightGBM classifiers for each behavioral tag. Our LightGBM classifier also used the same hyperparameters as the LightGBM benchmark provided with SOREL.

Each classifier was trained using five-fold cross validation, and mean ROC-AUC scores for each behavioral tag are listed in Tables \ref{tab:tag_ffnn_accuracy} and \ref{tab:tag_lgbm_accuracy}. 
The FFNN classifier achieved highest ROC-AUC scores for all 11 behavioral tags when using AVScan2Vec vectors. The LightGBM OvR classifier scored highest for nine tags when using AVScan2Vec vectors. Our results indicate that AVScan2Vec vectors encode behavioral traits that are present in AV scan data and can be used for effectively classifying malware. Furthermore, AVScan2Vec vectors yield a faster runtime and lower memory footprint than the other vectors due to its smaller size, while also offering marginal ROC-AUC score improvement on this task.

\begin{wraptable}[9]{R}{0.5\textwidth}
\vspace*{-10pt}

\centering
\resizebox{0.45\columnwidth}{!}{%
\begin{tabular}{@{}lrrrr@{}}
\toprule
 & AVScan2Vec & BWMD & EMBER & EMBER-N \\
\midrule
FFNN & \textbf{\textbf{80.49\%}} & 70.05\% & 23.22\% & 74.41\% \\
LightGBM & 71.47\% & 64.01\% & \textit{73.40\%} & 70.05\% \\
1-NN & \textit{65.43\%} & 48.85\% & 42.61\% & 58.95\% \\
\bottomrule
\end{tabular}
}
\vspace*{6pt}
\caption{Family classification accuracies. The highest overall accuracy is bolded. The highest accuracy per classifier is italicized. AVScan2Vec has the highest overall accuracy by over six percentage points, and performs either best or second-best with each classifier. }
\label{tab:family_accuracy}

\end{wraptable}

\subsubsection{Family Classification Results}
Next, we tested each vector's effectiveness when used for malware family classification. This is a challenging problem due to the large number of malware families in existence and the limited amount of labeled data available. Table \ref{tab:family_accuracy} lists the accuracy of FFNN,  LightGBM, and 1-Nearest Neighbor (1-NN) classifiers when trained using AVScan2Vec, BWMD, EMBER, and EMBER-N vectors. Each classifier was trained using the MOTIF dataset. The FFNN and LightGBM classifiers were trained using stratified five-fold cross validation, and the mean accuracy is shown. The LightGBM classifier used the hyperparameters from the LightGBM benchmark provided by the SOREL dataset, with minor adjustments which enable it to perform multiclass classification. Due to the significantly higher number of classes, the FFNN used a custom architecture rather than matching the FFNN benchmark from the SOREL dataset.

The 1-NN classifier's accuracy when trained using AVScan2Vec vectors indicates that related files tend to be embedded into nearby vectors with higher frequency than other approaches. When used to train a FFNN classifier, the AVScan2Vec vector achieved the highest overall accuracy by over six percentage points. It is clear that the AVScan2Vec vector format enables significant performance gains for malware family classification. We believe that EMBER's higher accuracy with the LightGBM classifier is due to its use of categorical features, which are better suited to tree-based models. However, AVScan2Vec yielded the second-highest performance with LightGBM and EMBER had the lowest accuracy with the other two classifiers by a large margin.

\subsubsection{Clustering Results}

Next, we evaluated the performance of each vector format when used for malware family clustering. We clustered the 2,962 MOTIF malware samples using K-Means and Hierarchical Agglomerative Clustering (HAC) with complete linkage. Both clustering algorithms used Euclidean distance and identified 323 clusters (the number of malware families in the dataset). Results are shown in Tables \ref{tab:kmeans_eval} and \ref{tab:hac_eval}. AVScan2Vec outperforms each of the other vector formats in homogeneity, completeness, and V-measure for both clustering algorithms. These results show that AVScan2Vec vectors have a stronger ability to group related malware samples than the other vector formats, while also grouping unrelated malware samples less frequently.

\vspace*{12pt}
\noindent
\resizebox{\columnwidth}{!}{%

\begin{minipage}[c]{0.48\textwidth}
\centering
\resizebox{\columnwidth}{!}{%
\begin{tabular}{@{}lrrrr@{}}
\toprule
 & AVScan2Vec & BWMD & EMBER & EMBER-N \\
\midrule
Homogeneity & \textbf{.770} & .713 & .683 & .749 \\
Completeness & \textbf{.826} & .677 & .666 & .764 \\
V-Measure & \textbf{.797} & .694 & .674 & .756 \\
\bottomrule
\end{tabular}
}
\vspace*{6pt}
\captionof{table}{K-Means family clustering evaluation results. AVScan2Vec vectors recieved the highest scores across all metrics.}
\label{tab:kmeans_eval}
\end{minipage}

\hspace*{32pt}

\begin{minipage}[c]{0.48\textwidth}
\centering
\resizebox{\columnwidth}{!}{%
\begin{tabular}{@{}lrrrr@{}}
\toprule
 & AVScan2Vec & BWMD & EMBER & EMBER-N \\
\midrule
Homogeneity & \textbf{.788} & .707 & .678 & .745 \\
Completeness & \textbf{.805} & .545 & .625 & .668 \\
V-Measure & \textbf{.797l} & .615 & .650 & .704 \\
\bottomrule
\end{tabular}
}
\vspace*{6pt}
\captionof{table}{Hierarchical Agglomerative Clustering evaluation results. Again, AVScan2Vec vectors scored highest on all metrics.}
\label{tab:hac_eval}
\end{minipage}
}

\newpage
\subsubsection{Nearest-Neighbor Lookup.}
Our final experiments 

\begin{wraptable}[11]{R}{0.5\textwidth}
\vspace*{-24pt}
\centering
\resizebox{0.45\columnwidth}{!}{%
\begin{tabular}{@{}lrrrr@{}}
\toprule
 & AVScan2Vec & BWMD & EMBER & EMBER-N \\
\midrule
Adware & \textbf{.914} & .799 & .883 & .889 \\
Flooder & .904 & .894 & .895 & \textbf{.906} \\
Ransomware & \textbf{.922} & .685 & .888 & .891 \\
Dropper & \textbf{.848} & .741 & .821 & .828 \\
Spyware & \textbf{.908} & .797 & .874 & .877 \\
Packed & \textbf{.924} & .817 & .906 & .908 \\
Crypto Miner & \textbf{.892} & .710 & .876 & .881 \\
File Infector & \textbf{.950} & .765 & .901 & .913 \\
Installer & \textbf{.865} & .691 & .846 & .857 \\
Worm & \textbf{.935} & .780 & .877 & .885 \\
Downloader & .888 & .788 & .867 & \textbf{.892} \\
\bottomrule
\end{tabular}
}
\vspace*{6pt}
\caption{10-nearest-neighbor F1 score per SOREL behavioral tag.} 
\label{tab:tag_knn_accuracy}

\end{wraptable}

\noindent assess AVScan2Vec's nearest-neighbor lookup performance. First, we randomly selected 10,000 files from SOREL-110000. Then, for each of these 10,000 files, we queried its 10-nearest neighbors from amongst the remaining 100,000 files. This was done for each of the vector formats. We computed per-tag F1 score by comparing the tags of the 10-nearest neighbors to the tags of the queried files.
Results are listed in Table \ref{tab:tag_knn_accuracy}. Using AVScan2Vec vectors for the 10-nearest lookup produced the highest F1 score for nine of the 11 behavioral tags, with significant improvement over BWMD and marginal improvement over EMBER and EMBER-N.

\begin{wraptable}[18]{R}{0.5\textwidth}
\vspace*{-12pt}
\centering
\resizebox{0.45\columnwidth}{!}{%
\begin{tabular}{@{}lrrr@{}}
\toprule
 & Precision & Recall & F1 Score \\
\midrule
Adware & .927 & .926 & .926 \\
Flooder & .923 & .928 & .925\\
Ransomware & .960 & .961 & .961 \\
Dropper & .931 & .927 & .929\\
Spyware & .956 & .954 & .955\\
Packed & .944 & .944 & .944\\
Crypto Miner & .936 & .940 & .938 \\
File Infector & .967 & .967 & .967\\
Installer & .916 & .915 & .915\\
Worm & .958 & .954 & .956\\
Downloader & .936 & .933 & .935\\

\bottomrule
\end{tabular}
}

\vspace*{6pt}
\caption{Evaluation results for 10-nearest-neighbor lookup using AVScan2Vec with DCI. The sub-linear runtime complexity resulted in an average query time of 16.41ms over 7,192,023 vectors.} 

\label{tab:tag_dci}
\end{wraptable}
    
\subsubsection{Dynamic Continuous Indexing.}

We are especially interested in how effectively AVScan2Vec can be used for nearest-neighbor lookup over massive dataset sizes. To test this, we employed Dynamic Continuous Indexing (DCI), a highly efficient nearest-neighbor lookup algorithm \cite{li2016fast}. Continuous indices were constructed for 7,192,023 AVScan2Vec vectors from SOREL in $\approx$22 minutes using default DCI hyperparameters and running on a single Intel Xeon Gold 6240R CPU. Then, the ten-nearest neighbors for each of the 100,000 remaining vectors were queried, with an average lookup time of 0.501 seconds per vector. DCI also supports multiprocessing, enabling very fast batch queries. Using 64 CPU cores enabled continuous indices to be constructed for the same dataset in 75 seconds and an average lookup time of 0.029 seconds per vector.
The sub-linear query runtime complexity of DCI would allow efficient AVScan2Vec vector lookups over even larger dataset sizes. Even in the worst case (linear scaling), repeating our prior experiment with 64 CPU cores over a dataset of 1 billion AVScan2Vec vectors would still allow an average 10-nearest neighbor lookup time of $\approx$4 seconds per vector. Precision, recall, and F1 score results for AVScan2Vec with DCI are shown per SOREL behavioral tag in Table \ref{tab:tag_dci}, showing that queries return relevant malware with high fidelity.

\section{Conclusion}
\label{sec:discussion}

We now address the limitations of our work, followed by our contributions and conclusions. AV scan data is very effective for tasks on all-malware datasets, but less so for mixed malware/benign datasets. The size of the AVScan2Vec model was limited due to hardware constraints. Adding additional layers to the Transformer encoder would almost certainly be beneficial. Due to concept drift caused by newly-emerging malware, AVScan2Vec would likely need to be re-trained monthly for optimal performance.

We overwhelmingly conclude that AV scan data contains informative and useful malware features. Furthermore, AV scan data is relatively inexpensive to obtain, even for production-scale malware corpora. Our AVScan2Vec model was able to successfully perform feature learning on AV scan data and embed malware into compact vector representations which can be used for downstream ML tasks. Our thorough validation demonstrated robust feature learning even under very restrictive conditions. AVScan2Vec was able to predict the correct token out of 10 million with over 78\% accuracy for reports with just a single positive detection.

We are preparing to publish all source code for our AVScan2Vec implementation, as well as scripts for training and evaluating it. These releases will enable AVScan2Vec to be trained on other datasets of AV scan reports, possibly with labels from other AV products or with future data. In addition, we will release the weights of the pre-trained and fine-tuned AVScan2Vec models. The pre-trained AVScan2Vec model can be easily adapted for fine-tuning on other custom tasks which use AV scan data as input. The fine-tuned model can be easily incorporated into a large-scale analysis pipeline and fits on a single GPU. The vectors predicted using the fine-tuned AVScan2Vec model can be used for related malware queries over enourous malware corpora and for various ML tasks.

We experimentally showed that AVScan2Vec offers marginal accuracy improvement for ML tasks using behavioral tags and significant improvement for tasks using family labels. AVScan2Vec's smaller vector size facilitates lower memory usage and faster vector comparisons. We believe that the combination of AVScan2Vec with DCI is especially potent. DCI's sub-linear query runtime complexity enables nearest-neighbor lookup over massive malware datasets, which has significant promise for aiding threat hunting efforts.

\bibliographystyle{ACM-Reference-Format}
\bibliography{sample-base}

\clearpage
\onecolumn

\appendix
\begin{table}[!tbh]
\section{Appendix}
\vspace*{0.5cm}

    \centering
        \caption{List of AV products included in the pre-training dataset. The accuracy of AVScan2Vec per AV product for Masked Token Prediction (M.T.P.) and Masked Label Prediction (M.L.P.) is also shown.}
        
        \vspace{-.75cm}\hspace{-.1cm}\includegraphics[width=0.96\columnwidth,keepaspectratio]{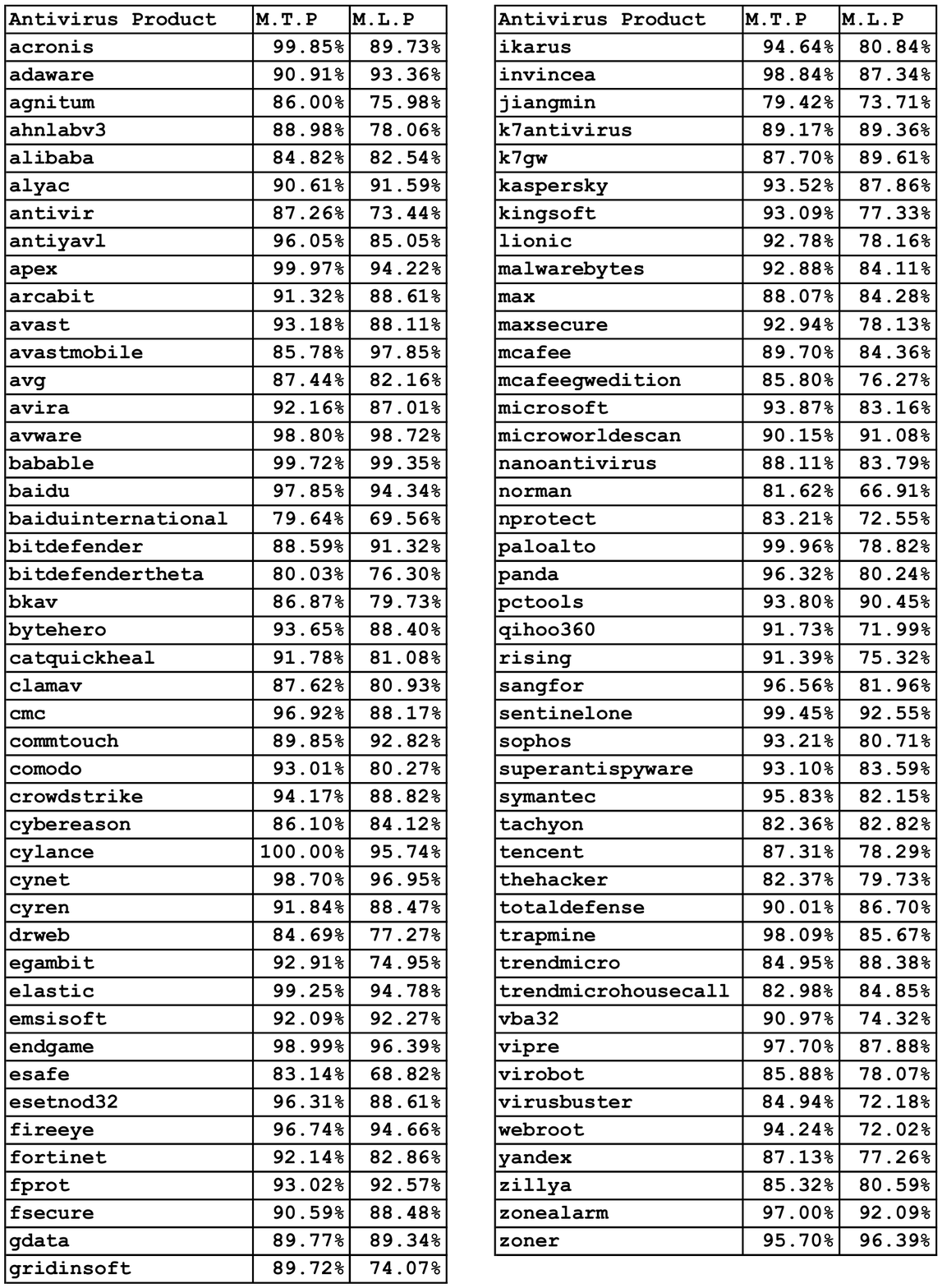}

    \label{tab:avscan2vec_av}
\end{table}

\end{document}